\documentstyle[preprint,prb,aps,epsfig]{revtex}
\begin{document}
\tightenlines
\draft


\title{Voronoi-Delaunay analysis of normal modes in a simple model
glass.}

\author{
V.A. Luchnikov$^{\dagger\ddagger}$, N.N. Medvedev$^{\dagger}$,
Yu.I. Naberukhin$^{\dagger}$ and H.R. Schober$^{\ddagger}$}

\address{{$^\dagger$} Institute of Chemical
Kinetics and Combustion,
630090 Novosibirsk, Russia}

\address{$^{\ddagger}$Theorie III, Institut f\"ur 
Festk\"orperforschung, Forschungszentum J\"ulich,
D-52425 J\"ulich, Germany}

\date{\today}

\maketitle

\begin{abstract}
We combine a conventional harmonic analysis of vibrations in a
one-atomic
model glass of soft spheres
with a Voronoi-Delaunay  geometrical analysis of the structure.  
``Structure potentials'' (tetragonality, sphericity or perfectness) 
are introduced to 
describe the shape of 
the local atomic configurations (Delaunay simplices) as function
of the atomic coordinates. Apart from the highest and lowest frequencies
the amplitude weighted ``structure potential''
varies only little with frequency. 
The movement of atoms in soft modes causes transitions
between different ``perfect'' realizations of local structure.
As for the potential energy a 
dynamic matrix can be defined for the  ``structure potential''. Its expectation
value with respect to the vibrational modes increases nearly linearly
with frequency and shows a clear indication of the boson peak.
The structure eigenvectors of this
dynamical matrix are  strongly correlated to the vibrational ones. Four
subgroups of modes can be distinguished.

\end{abstract}

\section{Introduction}
\hspace*{\parindent}
The  thermodynamic
properties  of  glasses at low temperatures differ
from   those   of   the
corresponding crystals \cite{phillips}. At low temperatures
the specific heat is strongly enhanced compared to the Debye contribution
stemming from the sound waves. The excitations underlying this enhancement
have been shown to be two-level systems below $T\approx 1$~K and nearly
harmonic vibrations above. The  vibrational density of state, $Z(\nu )$,
plotted as $Z(\nu )/\nu^2$ has a maximum, typically near 1~THz, the 
boson peak. 

This low temperature / low frequency behavior can be
described by the soft potential model. \cite{karpov:83,ilin:87}
In this model one assumes that one common type of structural unit is
responsible for the excess excitations. One introduces  an effective
potential describing the motion of this unit. Depending on the parameters
this potential is a single well or a double well. In the first case
it describes a low frequency localized vibration and in the second
tunneling through the barrier (two level systems) or relaxation over the
barrier. 
For low energies one can
give a general form for the distribution of the parameters
describing the effective potentials.
Fitting this model to the experimental data, one finds that 20--100
atoms or molecular units move collectively in the tunneling and in localized 
vibrations.\cite{BGGS,BGGPS} It should be emphasized that the concept
of low frequency localized vibrations is an idealization. These modes
will always interact with the sound waves of similar frequency
and, therefore, also among each other. This delocalises the modes 
and they are only quasi-local or resonant. 
Due to level repulsion, for sufficiently high densities
of these modes, the interaction will change their density of states
from $Z(\nu ) \propto \nu^4$ to $Z(\nu ) \propto \nu$ thus creating
the boson peak.\cite{GPPS} 
Such a model 
does, however, not say anything about
the physical nature of the localized modes or their origin
in different types of glass.

The problem of local dynamics in the amorphous state
is  closely  connected  with the problem of the so-called
medium-range order in glasses
\cite{elliott,malinovskii}.
Recently it has been shown \cite{struct,prb} that  a computer
model  of  amorphous
argon has  a  heterogeneous  structure containing regions 
of more ``perfect'' or  ``imperfect'' atomic arrangements
on a nanometer scale. In  the  regions  of perfect
structure  the  elementary packings of four neighboring atoms
(the Delaunay simplices) are close 
to either regular tetrahedra  or  quart-octahedra \cite{struct},
i.e. quarters of regular octahedra.
In  the
regions of imperfect structure the  local  configurations  of  the
neighboring atoms differ markedly from these ideal shapes. A
partial spectrum of the vibrational states of the atoms in the regions of
more ``imperfect''  structure   displays   an excess for
low-frequency modes.\cite{prb}

Quasi-localized
low-frequency vibrations have been observed in computer simulations
of the soft sphere glass (SSG)\cite{SL:91} and of numerous other materials,
such as e.g. SiO${}_2$ \cite{Jin93}, Se \cite{Olig93} 
in Ni-Zr \cite{Hafner94} and Pd-Si \cite{BalloneR95}, 
in amorphous ice \cite{Cho94} and in
amorphous and quasi-crystalline Al-Zn-Mg \cite{Hafner93}.
It was shown that these modes are centered at
atoms whose structural  surrounding
differs substantially from the average.
It  has  been  established  that  the
directions  of  the  eigenvectors  of  soft  vibrations  strongly
correlate   with   those    of the  relaxation    jumps    at low
temperatures.~\cite{SOL:93,OS:99}

One hypothesis on the origin of the soft mode is that the most active atoms 
oscillate between neighboring minima 
of the potential energy formed by a cage of surrounding atoms.~\cite{ziman}  
These minima correspond to 
some more ``perfect'' local arrangements of the atoms. The coupling to the
rest of the material changes this double well system to a soft single well one. 
One example for such a situation is the interstitial atom in an fcc metal.
A medium sized interstitial occupies the octahedral site. Increasing the
size of the interstitial atom the octahedral site becomes unstable and
the interstitial moves to an off-center position. The impeding instability
is indicated by low lying resonance vibrations \cite{dederichs:80,ehrhart:86}.
The instability in this example is caused by a local compression which
causes the simultaneous occurrence of high frequency localized vibrations.
In the glass the modes are more extended typically string like groups of some
twenty atoms
\cite{SO:96}. Instead of the single interstitial atom one has to take a
group of atoms and due to the laking symmetry the energy minima will be
shifted relative to each other. Keeping this in mind the underlying
mechanism can still be true. The simultaneous occurrence of low and
high frequency localized modes centered on one atom has indeed been
observed \cite{SL:91}.

In the present paper we want to verify and concretize
this notion for the SSG. For this purpose we combine 
the harmonic analysis of Ref.~\onlinecite{SL:91} with the 
Voronoi-Delaunay geometrical description of the local
structure used in Ref.~\onlinecite{prb}. 
First we shift
the atoms of the model along the eigenvector
of a low frequency quasi-localized normal mode and observe
the changes in the local atomic arrangements,
caused by the shifting.
This allows us to visualize
the specific transformations of local structure which accompany the 
movement of atoms in the soft vibrations. 
In a next step we calculate the atomic perfectness weighted with the
squared amplitudes of the vibrational modes. This quantity varies only
weakly with frequency. Considering that the vibrations are 
connected with changes in the geometry, we introduce a ``structural
dynamical matrix''. We will show in the following that there is 
a strong correlation between the ``structural eigenvectors'' and their
vibrational counterparts. This correlation divides the vibrations,
as regards structure changes, into separate classes: longitudinal and
transverse extended, high frequency localized and low frequency
quasi-localized modes.

\section{The soft sphere glass}

We use 55  glassy  configurations  of  500 atoms each, interacting 
via a soft  sphere
pair potential 
\begin{equation}
 u(r)=\epsilon \left(\frac{\sigma}{r} \right)^6 + 
       A\left( \frac{r}{\sigma} \right)^4 + B.
\end{equation}
To simplify the simulation the potential is cut off at $r/\sigma = 3.0$
and shifted by a polynomial with $A=2.54\times 10^{-5}\epsilon$ and
$B = -3.43\times 10^{-3}\epsilon$.
The calculations are done with a fixed atomic density, 
$\rho /\sigma^3 = 1$ and periodic boundary conditions. 
The configurations were obtained by a quench from the liquid
to $T=0$~K. From the pair correlation one finds a nearest neighbor
distance of around $1.1\sigma$. For more details  
see Ref.~\onlinecite{SL:91}.

The inverse sixth-power potential is a well-studied theoretical model
that mimics many of the structural and thermodynamic properties of
bcc forming melts including the existence, in its bcc crystal form,
of very soft shear modes.~\cite{hoover:71}
In the glassy structure one finds a boson peak with a maximum near 
$\nu = 0.1 (\epsilon / m \sigma^2 )^{1/2}$ extending to about
$\nu = 0.4 (\epsilon / m \sigma^2 )^{1/2}$. The enhancement 
of the vibrational density of states over
the Debye value is by a factor of 2.5.\cite{SO:96}.

As before the frequencies and eigenvectors of  normal vibrations are
calculated by the diagonalisation of the force constant  matrix.
Imaginary  frequencies are absent in the spectrum
because the system is in
an absolute local minimum of potential energy. 
For the given number of atoms the minimal $q$-value for sound waves
is $q_{\rm min} = 0.79 \sigma^{-1}$ giving minimal frequencies of
0.18 and 0.62 $(\epsilon / m \sigma^2 )^{1/2}$ for the transverse and
longitudinal sound waves, respectively. Resonant modes with frequencies
well below 0.18~$(\epsilon / m \sigma^2 )^{1/2}$
will, therefore, be seen as low frequency localized modes.
This is reflected in the participation ratios given in Ref.~\onlinecite{SL:91}.
One finds proper localized modes at frequencies 
$\nu > 2(\epsilon / m \sigma^2 )^{1/2}$ and 
(quasi-)localized low frequency modes with 
$\nu < 0.2(\epsilon / m \sigma^2 )^{1/2}$. The great majority
of modes ($0.2<\nu <2$) extends over the system. These latter modes have
been called diffusons \cite{fabian:96} due to their non-propagating
character. Nevertheless for the SSG as for other systems
it is possible to extract via the dynamic structure
factor some very broad ``phonon dispersions''.\cite{caprion:96}.

The SSG was used in extensive studies of the influence of quench rate
on the glass structure.~\cite{jund:97,jund:97a} In these studies the
Voronoi method was used to identify pentagonal rings which can be used
as signature of icosahedral packing.

\section{Voronoi-Delaunay description of local structure.}

By definition, the {\it Voronoi polyhedron} (VP)
of an atom is that region of space which is  
closer to the given atom than to any other 
atom of the system. A dual system spanning space is formed by 
the {\it Delaunay simplices} (DS).
These are  tetrahedra formed by four atoms which
lie on the 
surface of a sphere which does not contain
any other atom. 
Both VP and DS fill the space of the system without 
gaps and overlaps. In our calculations we do a Voronoi-Delaunay tessellation
of the glass configurations by the algorithm described in 
Ref.~\onlinecite{VD_tesselation}.

It was found earlier that two main types of DS are predominant
in mono-atomic glasses \cite{DS_classes_1,DS_classes_2}
namely DS similar to ideal tetrahedra and DS resembling a quarter
of a regular octahedron (quart-octahedron).  
Following Ref.~\onlinecite{T_O_definition} 
we introduce as quantitative measure of
{\it tetragonality} of a DS
\begin{equation}
\label{eq_tetr}
T=\sum_{ i<j } \frac{(l_i-l_j)^2}{15\overline{l}^2}
\end{equation}
where $i$ and $j$ designate the edges of the simplex,
and $\overline{l}$ is the average edge-length.
This measure was constructed to be
zero for ideal tetrahedron and to increase
with distortion. 
For computational reasons we slightly modify the previous measure of 
{\it octagonality} using:
\begin{equation}
O=\left\{ \sum_{m=1}^{6}g_{m}O_{m}^{-1}\right\}^{-1}
\label{oct_weighted}
\end{equation} 
\noindent where
$$g_m=\frac{e^{3\frac{{\delta}_{m}}{\sigma}}}
{\sum_{i=1}^{6}e^{3\frac{{\delta}_{i}}{\sigma}}},\hspace{2mm} 
{\delta}_{m}=\frac{l_{m}-\overline{l}}{\overline{l}},\hspace{2mm} 
\sigma = \left[\frac{1}{6}\sum {\delta}_m^2\right]^{\frac{1}{2}},
$$
\noindent and
\begin{equation}
O_m=\sum_{i<j;\hspace{1mm} i,j\neq m }
\frac{(l_i-l_j)^2}{10\overline{l}^2}
+\sum_{i\neq m} \frac{(l_i-{l_m}/\sqrt{2})^2}{5\overline{l}^2}
\label{oct_marked}
\end{equation}
In a perfect quart-octahedral DS one edge
is $\sqrt{2}$ times larger
then the other edges. In Eq.~(\ref{oct_marked}), 
the previously used measure \cite{T_O_definition}, it was originally assumed
that the $m$-th edge is the longest.
The modified expression  (\ref{oct_weighted}) 
weights the six 
possible values $O_m$ in such a way that the smallest one dominates.
The octagonality thus tends to zero when the DS is close to the 
perfect quart-octahedron.
This weighting allows us to avoid the use of logical functions
for the selection of the maximal edge and guarantees 
differentiability 
which is essential for our investigation.
For the relevant  low values of $O$, i.e simplices close to
quart-octahedral shape, our expression reproduces the values of the
original definition.

The tetrahedral and 
quart-octahedral DS can be unified in one class of 
``perfect'', or ``ideal'' simplices \cite{struct}. 
We measure the {\it ideality} of the DS shape by
\begin{equation}
\label{eq_id}
S=\left[
g_T\left(\frac{T}{T_c}\right)^{-1}+
g_O\left(\frac{O}{O_c}\right)^{-1}\right]^{-1}
\end{equation}
where
$$
g_T=\frac{e^{-3\frac{T}{T_c}}}{e^{-3\frac{T}{T_c}}+e^{-3\frac{O}{O_c}}},
\hspace{3mm}
g_O=\frac{e^{-3\frac{O}{O_c}}}{e^{-3\frac{T}{T_c}}+e^{-3\frac{O}{O_c}}}.
$$
$S$ tends  to zero when the simplex takes the 
shape of an ideal tetrahedron {\it or} quart-octahedron.
Contrary to the expression proposed in Ref.~\onlinecite{struct}
our measure is differentiable with respect to the atomic coordinates. 
The relative weights 
of tetrahedricity and octahedricity, $T_c=0.016$, $O_c=0.033$,
are taken from Ref.~\onlinecite{T_O_definition}.
The relation  of the values $T$, $O$ to the 
distortion of a DS  can be also seen from the 
values $T_O \simeq 0.050$ of the  tetrahedricity 
of an ideal quart-octahedron and $O_T \simeq 0.084$
octahedricity of an ideal tetrahedron.

Each atom in the glass is corner of approximately 24 DS. 
The structural environment
of an individual atom can be characterized by the 
average ideality of over these DS
\cite{prb}:
\begin{equation}
     S_{\rm atom} = \frac{1}{n_{DS}}\sum_{i=1}^{n_{DS}} S_{i}
\label{eq_satomic}
\end{equation}
where $n_{DS}$ is the number of DS surrounding the atom.

Another widely used measure of the atomic
neighborhood is the {\it sphericity} of the Voronoi cell:    
\begin{equation}
\label{eq_sph}
   Sph = \frac{1}{36\pi}\frac{F^3}{V^2} - 1.  
\end{equation}
Here $F$ is the surface area of the VP, and $V$ is its volume.
This measure is minimal for a sphere , $Sph=0$,
and again increases with distortion.  
 
In analogy to the potential energy one can take the total
tetrahedricity, ideality or sphericity to characterize the structure.
We introduce an average ``structure potential'' by
\begin{equation}
\langle T \rangle =  \frac{1}{N_{DS}}\sum_{i=1}^{N_{DS}}T_i
\label{eq_tav}
\end{equation}
and analogously $\langle S \rangle$ and $\langle Sph \rangle$ 
where $N_{DS}$ is the number of DS in the system.
Since dynamics is concerned with the motion of the atoms it is often
more useful to average over the atomic quantities defined by 
Eq.~\ref{eq_satomic}
\begin{equation}
\langle T_{\rm atomic} \rangle =  \frac{1}{N}\sum_{i=1}^{N}T_{\rm atomic} .
\label{eq_avtatomic}
\end{equation}
Both definition give similar values.

In table~\ref{table1} we compare the values of the three measures for an
ideal fcc-structure an icosahedron and our glass. The values of the glassy
structure clearly deviate from the ones of both the ideal configurations.
It is, however, not possible to define unambiguously a nearness to either
structure using these measures. 

\section{Soft vibrations and  
change of the local structure}

\begin{figure}
\epsfig{figure=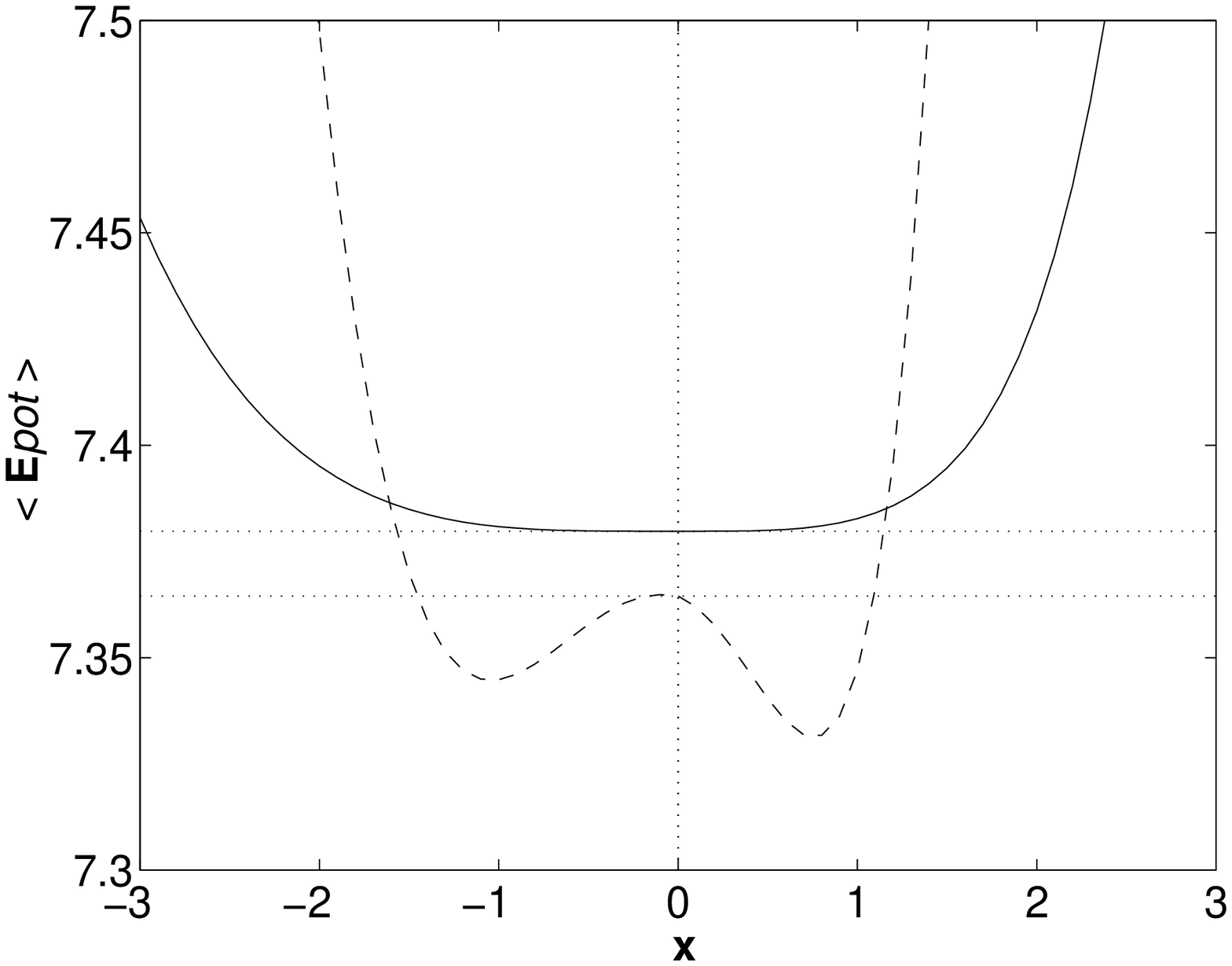,width=10cm,angle=0}
\caption{
Potential energy of atoms as function of the normal 
coordinate, $x$, corresponding to a soft mode.
{\it Solid line:} average potential energy of all atoms
in the system. $\langle E_{\rm pot}(x=0) \rangle \approx 7.388\epsilon$.
{\it Dashed line:} average potential energy of the 13 atoms most active
in the soft mode.
$\langle E_{\rm pot}(x=0) \rangle_{13} \approx 7.364\epsilon$. 
}
\label{u(x)}
\end{figure}

Using the quantities defined above we will now 
illustrate for one example of a quasi-localized soft mode the
relationship between softness and local geometry.
In Fig.~\ref{u(x)} (solid line) we show the average potential energy per atom
as function of
the displacement along a single soft eigenmode, i.e. one of the soft potentials
which are described by the soft potential model\cite{karpov:83,ilin:87} 
discussed in the introduction.
The atoms are shifted along the direction of the $3N$-dimensional
eigenvector ${\bf e}$ as
\begin{equation}
{\bf R}^n(x)={\bf R}_0^n + x{\bf e}^n.
\label{shift}
\end{equation}
Here ${\bf R}_0^n$ is the equilibrium position of atom $n$. For simplicity
we have not normalized the amplitude x to an effective atomic amplitude as
is usually done in the soft potential model.   
Fig.~\ref{u(x)} corresponds to a very well localized soft mode
with $\nu = 0.04 (\epsilon / m \sigma^2 )^{1/2}$,
effective mass $13 m$ and participation ratio $0.14$. These values
would guarantee a very narrow resonance in the infinite 
medium \cite{dederichs:80}.

As mentioned in the introduction it has been speculated that the soft modes 
in glasses originate  in some ``soft'' atomic configurations where,
in the extreme case,
the atoms are stabilized by the embedding matrix  in 
a position lying between minima
of the potential energy given by its near neighbors.
In Fig.~\ref{u(x)} we show by the dashed
line the average potential energy of the 13 most active atoms
of our soft mode, i.e. the atoms with the largest amplitude ${\bf e}^n{\bf e}^n$.
This {\it partial} potential energy is
indeed double-well shaped with  minima 
at $x_m \approx\pm 1.0$ which 
corresponds to maximal displacements of 
individual atoms by $|{\bf R}^n -{\bf R}_0^n|\approx 0.27\sigma $ 
from the equilibrium configuration.   
Note that at $x=0$ the selected atoms have somewhat smaller 
potential energy
than the  average. This reflects the reduced number of nearest neighbors 
reported earlier.\cite{SL:91}

\begin{figure}
\epsfig{figure=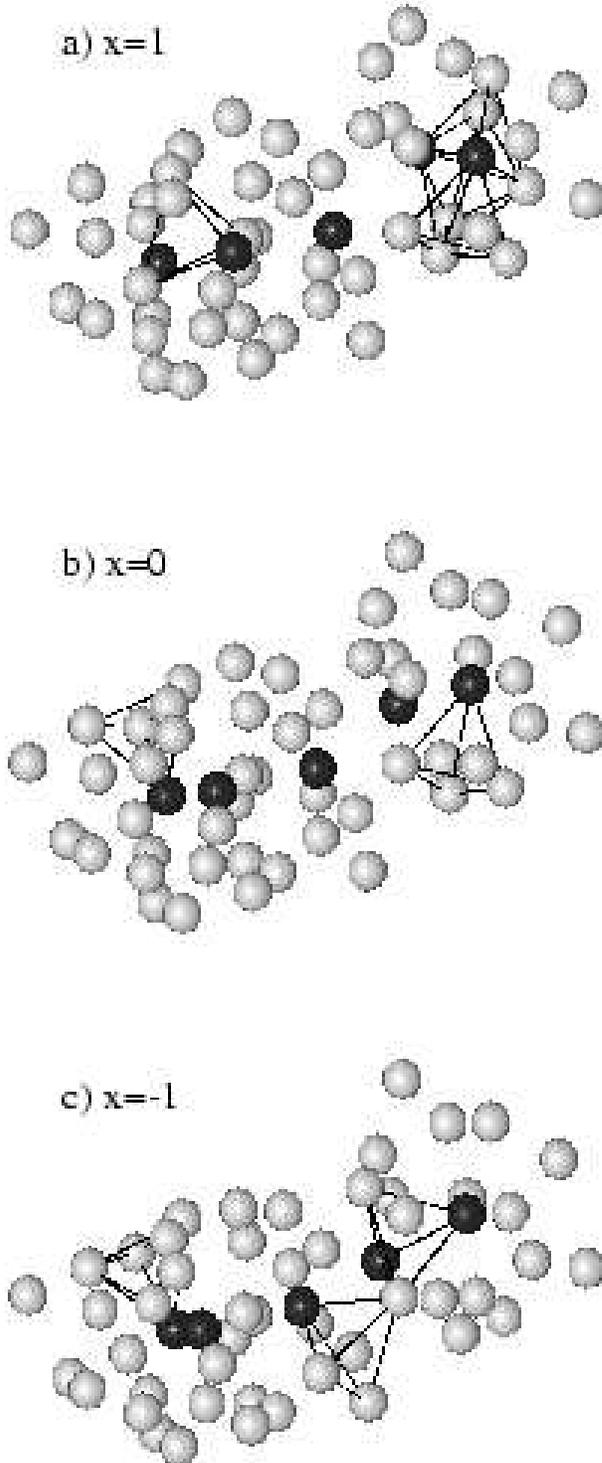,width=8cm,angle=0}
\caption{
Variation of local structure in the soft mode. 
{\it Black spheres:} the five atoms most active in the 
mode. {\it Gray spheres:} geometrical neighbors of the 
active atoms. DS of perfect tetragonal shape    
($T < 0.003$) are shown. {\it a)} 7 DS,
{\it b)} 2 DS, {\it c)} 4 DS. 
}
\label{tetr_trans}
\end{figure}

\begin{figure}
\epsfig{figure=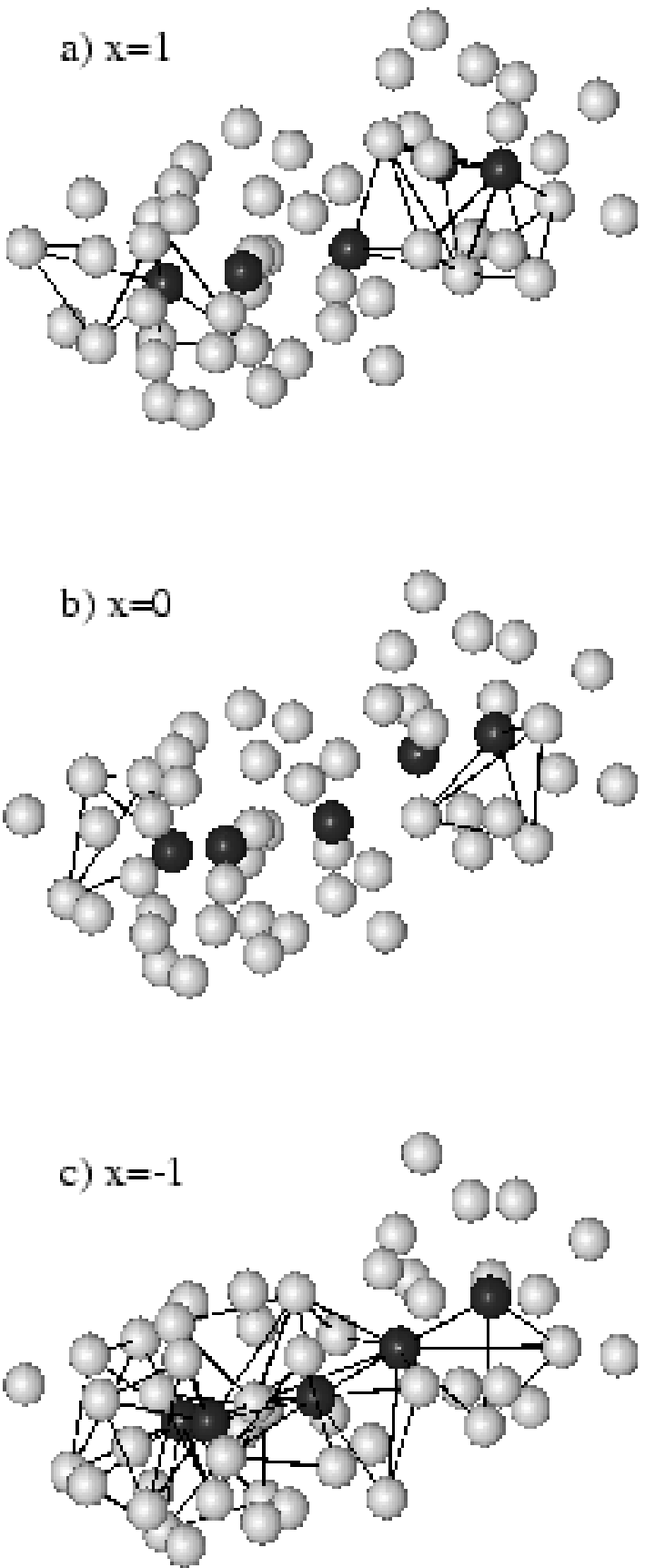,width=8cm,angle=0}
\caption{
Similar to Fig.~\ref{tetr_trans}: DS of perfect
quart-octahedral shape ($O < 0.008$).
{\it a)} 5 DS,
{\it b)} 2 DS, {\it c)} 12 DS.
}
\label{octa_trans}
\end{figure}

In order to understand the changes in the local structure 
as the atoms oscillate between the two partial minima of the potential
energy, we look at the changes of the Delaunay tesselation
caused by the displacements of the atoms in the normal mode.  
In Figs.~\ref{tetr_trans} and \ref{octa_trans} the five most 
active atoms of the soft mode are shown by the black spheres,
and their geometrical neighbors by  gray spheres.
We consider two atoms as geometrical neighbors
if they share a DS. The DS are visualized 
by line segments.

In Fig.~\ref{tetr_trans} we concentrate on DS with nearly ideal tetrahedral
shape. For the sake of clarity only the most perfect tetrahedra ($T < 0.003$)
are drawn. In equilibrium
($x=0$) two tetrahedra are found which satisfy this
condition (Fig.~\ref{tetr_trans},~b).
After shifting in the ``positive''
direction (Fig.~\ref{tetr_trans},a) one perfect tetrahedron has
disappeared, but 6 new ones appeared.
In particular, a 5-fold ring of perfect tetrahedra
is created (seen sideways in the right half of the picture).
This ring is known to be the densest possible packing
of 7 equal spheres. 
After the displacement in the ``negative''
direction (Fig.~\ref{tetr_trans},c) again one
tetrahedron is lost, but 4 new perfectly
tetrahedral DS are gained.

Similarly Fig.~\ref{octa_trans} demonstrates 
the appearance of new quart-octahedral DS in the neighborhood
of the active atoms. Only DS with $O < 0.008 $ are shown.
For $x=1$ one perfect octahedron, 
consisting of 3 active
atoms and 3 of their neighbors, is observed. Together
with the 5-fold ring of the perfect tetrahedra, 
it forms locally a pattern of perfect, 
non-crystalline structure which does not exist at $x=0$.
At $x=-1$, a large dense cluster of 12 ideal quart-octahedra 
appears (Fig.~\ref{octa_trans},~c). It indicates 
several octahedral configurations in the neighborhood
of the active atoms, although they cannot be seen clearly
on the figure.     

\begin{figure}
\epsfig{figure=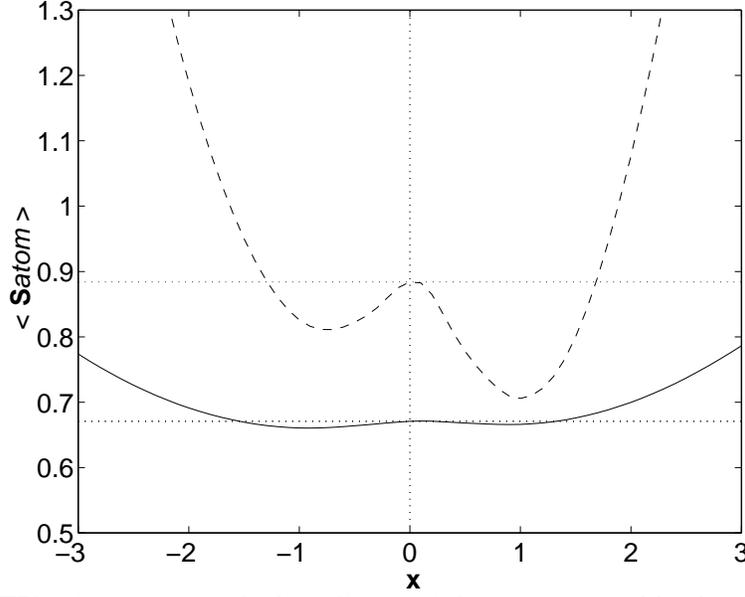,width=10cm,angle=0}
\caption{
Average ideality, $S_{\rm atom}$, of the nearest neighborhood as the function
of the normal coordinate, $x$, of the soft mode used in Fig.~(\ref{u(x)}).
{\it Solid line:} $S_{\rm atom}$ averaged over all atoms of the system.
$\langle S_{\rm atom}(x=0) \rangle \approx 0.67$.  
{\it Dashed line:} $S_{\rm atom}$ averaged over 13 active atoms.
$\langle S_{\rm atom}(x=0) \rangle_{13} \approx 0.88$. 
}
\label{sa(x)}
\end{figure}

In general, the number of perfect DS
(tetrahedra or quart-octahedra) increases as the atoms 
are shifted from the equilibrium position to the local minima 
of the partial potential energy. This tendency is summarized in the
double-well behavior of the average ideality,
$\langle S_{\rm atom} \rangle $,
as function of the normal coordinate of the mode
(Fig.~\ref{sa(x)}, dashed line) \cite{plot}.
We remind that a lower values of $S_{\rm atom}$ means a more perfect
atomic neighborhood.
Note that the minima of the partial ideality are situated
approximately at the same values of $x$ as the minima
of the partial potential energy. 
In the equilibrium position ($x=0$) the active atoms
have relatively imperfect neighborhoods compared to
the rest of the atoms. 
After the displacements this is considerably ``improved''.  

The curve of the average perfectness
$\langle S_{\rm atom}(x)\rangle$, averaged over all atoms (solid line)
is almost flat at $x=0$ and resembles the behavior of
the average potential per atom energy of the mode
(Fig.~\ref{u(x)}, solid line).

The double-well behavior of the partial potential 
energy and the ideality of the atomic environment 
is specific for a number of low-frequent vibrations.
It becomes less pronounced as the frequency increases. 
However, we have not noticed a sharp transition between
localized and delocalized low-frequency modes.   
Displacements of the atoms along the
modes with medium and high frequencies 
also destroy DS of near perfect shape.  

The geometrical peculiarities
of the random soft sphere packing play an important role
for the localized low-frequent vibrations. 
These vibrations can be visualized as complex collective
motions which organize the atoms in some ``perfect'' but
non-crystalline arrangements. Although these arrangements consist
of elements present in the crystalline structure (tetrahedra and
quart-octahedra) these are connected to each other in a way
which is incompatible with rotational and translational symmetries
of the crystal. Slight  deviations of the shape of the DS make
a variety of spatial arrangements possible which differ from the
closed packed fcc and hcp ones. The pentagonal rings typical for
locally icosahedric structure is one such example, compare e.g. 
Ref.~\onlinecite{jund:97}. The lack of translational symmetry
restricts these structures to ranges of a few interatomic distances.

\section{Correlation between structure and vibration}

We have seen that a low frequency quasi-localized vibration has a specific
impact on the structure surrounding  the most active atoms of the vibration.
We will now investigate how far a general relationship between structural
measures and dynamics can be seen. We will here concentrate on tetragonality,
Eq.~\ref{eq_tetr}.
Qualitatively we find the same trends also for ideality, Eq.~\ref{eq_id},
and sphericity, Eq.~\ref{eq_sph}. 

\begin{figure}
\epsfig{figure=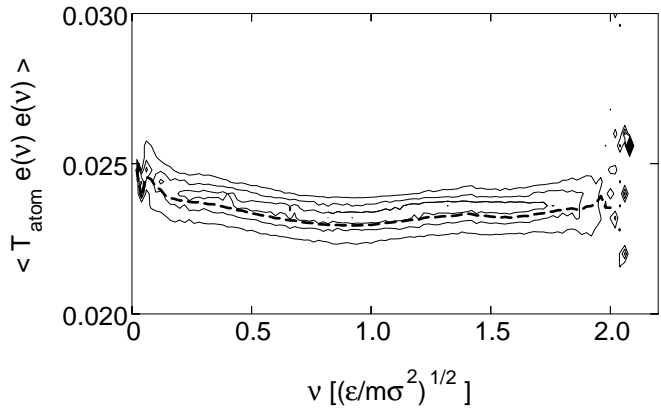,width=10cm,angle=0}
\caption{Amplitude weighted atomic tetragonality as function of vibrational
frequency, mean value (dashed line) and equidistant contour lines.}
\label{fig_tee}
\end{figure}

As a first possible relation between structural measures and vibration one
can take the atomic tetragonality weighted by the amplitudes on the atoms.
This would show whether e.g. atoms with low values of $T_{\rm atom}$
participate particularly strongly in vibrations in some frequency range.
We define
\begin{equation}
T(\nu ) = \langle \frac{1}{N} 
        \sum_n T_{\rm atomic}^n {\bf e}^n(\nu ) {\bf e}^n(\nu ) 
          \rangle
\label{eq_tee}
\end{equation}
where ${\bf e}^n(\nu )$ stands for the three components 
on atom $n$ of a vibrational eigenvector to frequency $\nu$ and 
$\langle \dots  \rangle $ denotes averaging over configurations and eigenvectors
to similar frequencies. Taking the average value, the dashed line in
Fig.~\ref{fig_tee}, one observes only a slight variation with frequency. Only
at the smallest frequencies a small upturn is found. The contour plot shows
that the $T(\nu )$-values fall in general into a  narrow band. This is 
different for the high frequency localized modes ($\nu > 2 (\epsilon /
m\sigma )^{1/2}$). These modes have a large spread of  $T(\nu )$-values
without a clear preference to large or low atomic tetragonalities.
This large spread is a direct consequence of the strong localization to
even single atoms. The low frequency modes involve always larger numbers
of atoms, from 10 upwards, and therefore average over many different 
atomic tetragonalities. 
 
In the previous section we noted a connection between low frequency
vibrations and changes of structural elements. In order to quantify
this notion we will now treat the average tetragonality, Eq.~\ref{eq_tav}, 
as a structural potential and in analogy to the usual dynamic matrix
define a {\it tetragonality matrix}
\begin{equation}
\label{eq_tetmat}
{\cal T}_{\alpha \beta}^{m n} = \frac{\partial^2 \langle T \rangle}
        {\partial R^m_\alpha \partial R^n_\beta} .
\end{equation}
Diagonalisation of this matrix gives the eigenmodes of tetragonality change
and the corresponding eigenvalues, which we will denote by ${\bf e}_T$ and
$\lambda_T$, respectively. To keep in line with the vibrations we use
a tetragonality frequency $\nu_T = \sqrt{\lambda}/{2\pi}$.

\begin{figure}
\epsfig{figure=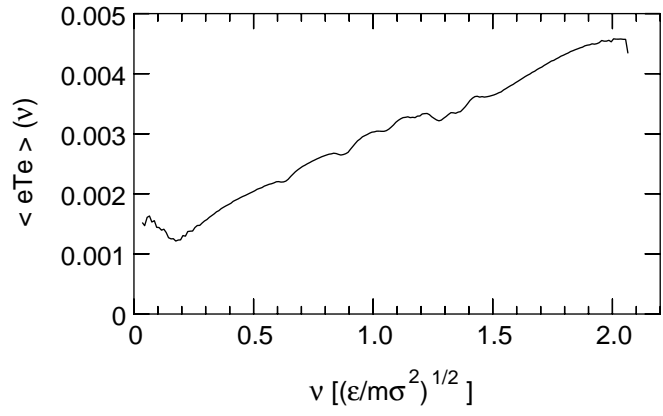,width=10cm,angle=0}
\caption{Expectation value of tetragonality with respect to vibration
modes versus frequency.}
\label{fig_ete}
\end{figure}

In analogy to Eq.~\ref{eq_tee}, where we defined an amplitude weighted
tetragonality as function of frequency,
we calculate the expectation value of
the tetragonality matrix with respect to the vibrations, i.e. an amplitude 
weighted structural curvature,
\begin{equation}
\langle e(\nu ) {\cal T} e(\nu ) \rangle =
      \left\langle \sum_{\alpha \beta}^{m n}    
       e_\alpha^m (\nu ) {\cal T}_{\alpha \beta}^{m n} e_\beta^n (\nu )
       \right\rangle .
\label{eq_ete}
\end{equation}
This expectation value shows several interesting features, Fig.~\ref{fig_ete}.
Most obviously there is a clear more or less linear increase with
frequency. This linearity breaks down at the lowest frequencies $\nu <
0.2 (  \epsilon /
m\sigma )^{1/2}$), i.e. in the frequency range of the boson peak, where
we find a distinct upturn. This upturn corresponds to the one in 
Fig.~\ref{fig_tee}
but is much more pronounced. It clearly indicates a structural difference
of the excess modes in the boson peak. 
where a small
maximum resembling a boson peak is seen. It should be remembered that
the translational invariance requires $\langle e(\nu ) {\cal T} e(\nu ) \rangle
\to 0$ for $\nu \to 0$. For pure translation we get of course zero. Due to
the limited system size
sound waves below $\nu < 0.2 (\epsilon /
m\sigma )^{1/2}$ were eliminated and we do not see the increase 
towards this ``structural boson peak'' on the low frequency side.
The small dips of the curve for
$\nu \approx 0.62, 0.88, ....(\epsilon /m\sigma )^{1/2}$ 
coincide with the frequencies
of the longitudinal sound waves in the SSG.\cite{S:99}

\begin{figure}
\epsfig{figure=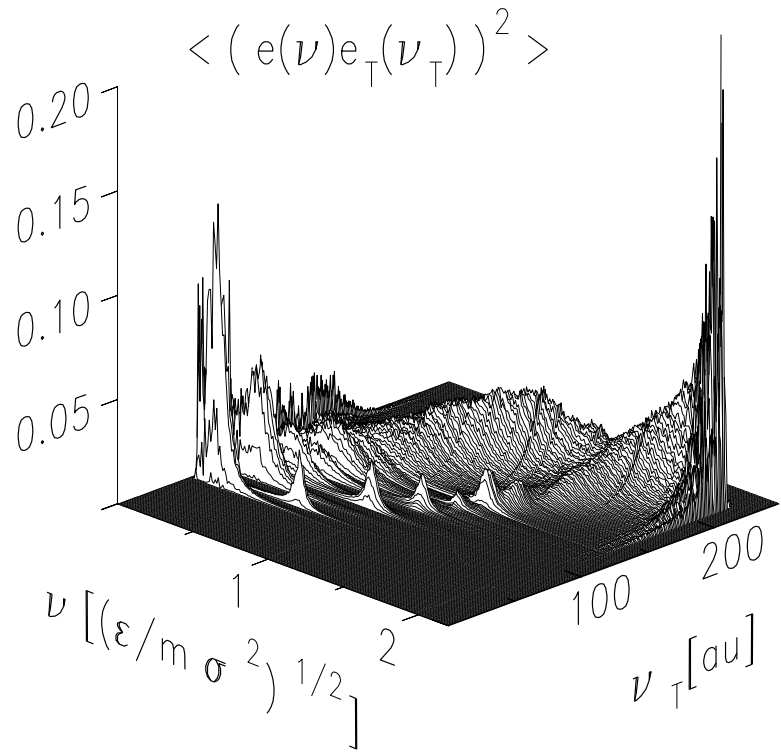,width=15cm,angle=0}
\caption{Correlation of vibrational eigenvectors, ${\bf e}$, and the 
eigenvectors, ${\bf e_{\rm T}}$, 
of the tetragonality matrix  as function of vibrational
frequency $\nu$ and tetragonality frequency $\nu_{\rm T}$.}
\label{fig_eeT}
\end{figure}

To get some deeper insight into the interplay of vibration and structure
change we calculate the correlation matrix between the vibrational 
eigenmodes and their tetragonality counterparts
\begin{equation}
\langle e(\nu ) e_T(\nu_T) \rangle
   = \left\langle \sum_{n \alpha} \left( e^n_\alpha(\nu ) {e_T}^n_\alpha(\nu_T )  
     \right)^2 \right\rangle .
\label{eq_eeT}
\end{equation} 
The resulting correlation, Fig.~\ref{fig_eeT}, shows several interesting
features. First there is a clear overall correlation as expected
from Fig.~\ref{fig_ete}. The correlation is highest for the highest 
frequency modes. From the participation ratios\cite{SL:91} on can see
that both vibrational and tetragonality modes are localized for the
highest frequencies. For the great majority of modes two groups can be
distinguished. The largest contribution stems from a broad band stretching
from the lowest $\nu$-values to the peak at the maximal $\nu$-values.
In front of this band (higher $\nu$-values) there is a smaller one which
can be identified as being due to longitudinal phonons\cite{S:99} which
are well separated from the other vibrations in the SSG. A third group
is seen as narrow ridge at low $\nu$ covering a major part of the
$\nu_T$ range. This last feature shows again the difference between the
quasi-localized low frequency modes and the rest of the spectrum. In a
larger system interaction will of course mix these features. This does,
however, not change the underlying nature of the ``naked'' modes.\cite{SO:96}
Fig.~\ref{fig_eeT} does not only show separate peaks for the longitudinal
phonons permitted by the system size but also regarding the $\nu_T$-direction
separate ``phonons'' are seen, both transversal and longitudinal. Checking
the participation ratios of $e_T$ one finds all modes with low $\nu_T$
to be extended, no low frequency localized modes are seen. The
observed correlation is insufficient to predict localization at low
frequencies. This is not too surprising as it has been observed earlier
\cite{SL:91} that these modes are produced by a subtle interplay of
local compression and 
in addition a resulting soft direction in configurational space
involving several atoms. The tetragonality reproduces the first feature,
seen in the high frequency modes, but not the second one. 

\begin{figure}
\epsfig{figure=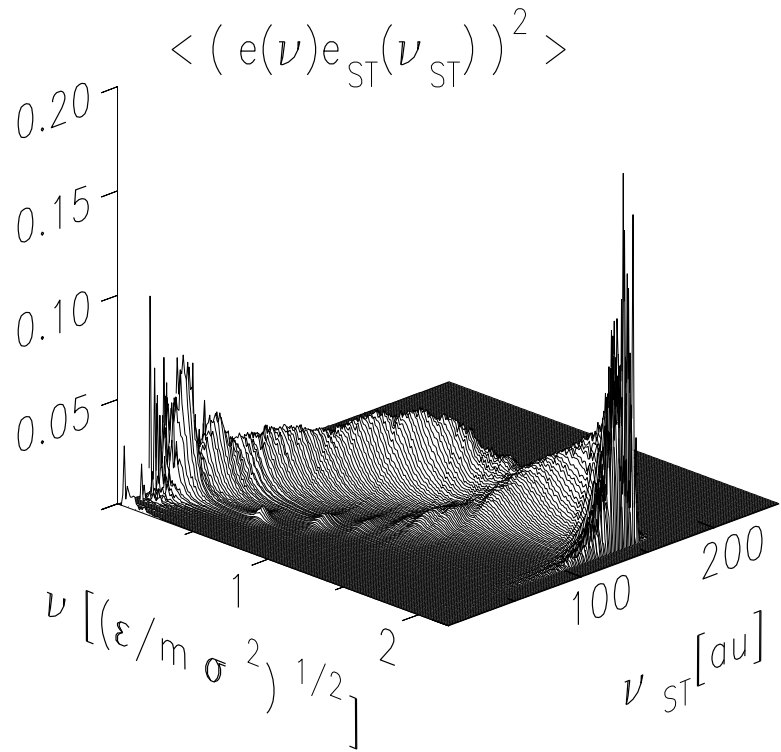,width=15cm,angle=0}
\caption{Correlation as in Fig.~\ref{fig_eeT} for a mixed measure of 
tetragonality and ideality, Eq.~\ref{eq_ST}.} 
\label{fig_eeST}
\end{figure}

To illustrate the fine details governing localization on one hand, and
the stability of the overall correlation on the other one we repeat the above 
calculation for a mixed measure of ideality and tetragonality
\begin{equation}
\label{eq_ST}
{\cal ST}_{\alpha \beta}^{m n}  = \frac{0.6}{{\rm tr}{\cal T}}
                                   {\cal T}_{\alpha \beta}^{m n} +
                                  \frac{0.4}{{\rm tr}{\cal S}}
                                   {\cal S}_{\alpha \beta}^{m n}.   
\end{equation}
The weighting of ${\cal T}$ and ${\cal S}$ was chosen somewhat arbitrarily to 
move the lowest eigenvalues to $(2\pi \nu_{ST})^2 \approx 0$.
Qualitatively the correlation, Fig.~\ref{fig_eeST} is the same as the
one for tetragonality, Fig.~\ref{fig_eeT}. 
The phonons in
$ST$-space are no longer so clearly discernible but  low frequency 
localized $ST$-modes are found which are correlated to the low 
frequency vibrations. The occurrence of the low frequency modes for the
mixed ${\cal ST}$ matrix is in agreement with Fig.~\ref{u(x)}
where the near zero curvature of $S_{\rm atom}$ is due to
changes of both tetrahedricity and octahedricity.
The difference between Figs.~\ref{fig_eeT} and
\ref{fig_eeST} illustrates that the formation of low frequency quasi-localized 
modes depends much more subtly on structural details than is the case for 
high frequency modes.

\section{Conclusion}
We have shown that the
Voronoi-Delaunay geometrical approach gives an insight into
the geometrical effects underlying the vibrations in the glass. The lowest
frequency quasi-localized vibrations can be envisaged as being
caused by an instability of the local geometry which is stabilized
by the embedding lattice. A group of atoms is trapped between
two configurations which can be considered as more perfect. We 
introduce different measures to quantify this perfectness.
For the great majority of modes there is only a weak correlation
between the amplitude on the single atoms and their perfectness.
This reflects the delocalisation of the modes. The high frequency
localized modes which are concentrated on one or two atoms show
a large scatter of their geometrical parameters which indicates
that they are caused by different local distortions. At the low
frequency side there is a small increase of tetragonality which
is, however, masked by the width of the distribution.

Introducing structural dynamic matrices correlation effects are 
clearly observable. These correlations divide the vibrations into
different groups. First there are two bands of extended modes, 
longitudinal and transverse ones. The separation of these two band is due to 
the large difference in longitudinal and transverse sound velocity for
the considered model. At high frequencies localized vibrations are
correlated to high frequency structural modes. At the lowest frequencies,
the region of the boson peak, the vibrations show a 
distinctly different correlation behavior. This is a clear indication of 
their structural origin. Using a suitable mixture of structural measures
low frequency structural modes can be defined which are correlated to
the low frequency quasi-localized (resonant) modes.

\section{Acknowledgment}
One of the authors (V.A.L.) gratefully acknowledges the hospitality and 
financial support of the Forschungszentrum J\"ulich.

\onecolumn
\onecolumn

\begin{table}
\caption{Values of ideality, tetragonality and sphericity 
for an atom in an ideal fcc-lattice, at the center of an icosahedron 
and for the glass, averaged according to
Eq.~\ref{eq_avtatomic}. }
\begin{tabular}{lccc}
     & $\langle S_{\rm atom} \rangle$ &
       $\langle T_{\rm atom} \rangle$ &
       $\langle Sph_{\rm atom} \rangle$\\
fcc lattice    & 0.0 & 0.039& 0.346\\
icosahedron    & 0.095 & 0.0015 & 0.325\\
SS-glass       & 0.66 $\pm$ 0.17& 0.023 $\pm$ 0.005& 0.358 $\pm$ 0.025\\
\end{tabular}
\label{table1}
\end{table}

\begin{thebibliography}{99}
\bibitem{plot}
To obtain this plot, we do not make a new tesselation for 
each value of $x$. The tetragonality and octahedricity 
are calculated for 
the quartets of atoms forming the DS at $x=0$.
This does not introduce large errors in the value of
$\langle S_{\rm atom} \rangle $ because 
for small displacements, the majority of atoms
would be grouped in the same DS if a new
tesselation is made.
\end{thebibliography}
\end{document}